\documentclass{webofc}
\usepackage[varg]{txfonts}   % Web of Conferences font

\usepackage{bm}
\usepackage{hyperref}       % hyperlinks
\usepackage{url}            % simple URL typesetting
\usepackage{booktabs}       % professional-quality tables
\usepackage{amsfonts}       % blackboard math symbols
\usepackage{nicefrac}       % compact symbols for 1/2, etc.
\usepackage{microtype}      % microtypography

\usepackage{graphicx}
\graphicspath{{figures/}}
\usepackage{physics}
\usepackage{subcaption}
\usepackage{wrapfig}

\usepackage{tikz}
\usetikzlibrary{quantikz}
\usepackage{bm} %Bold math  symbol
\usetikzlibrary{shapes,arrows}
\usetikzlibrary{positioning}
\usetikzlibrary{shadows}

\newcommand{\figref}[1]{Fig.~\ref{#1}}

\newcommand{\secref}[1]{Section~\ref{#1}}
\newcommand{\eqnref}[1]{Eq.(\ref{#1})}
\usepackage[en-GB]{datetime2}

\begin{document}
\title{Dual-Parameterized Quantum Circuit GAN Model in High Energy Physics}

\author{\firstname{Su Yeon} \lastname{Chang}\inst{1,2}\fnsep\thanks{\email{su.yeon.chang@cern.ch}} \and
        \firstname{Steven} \lastname{Herbert}\inst{3, 4}\fnsep\thanks{\email{steven.herbert@cambridgequantum.com}} \and
        \firstname{Sofia} \lastname{Vallecorsa}\inst{1}\fnsep\thanks{\email{sofia.vallecorsa@cern.ch}} \and
       \firstname{El\'{\i}as F.} \lastname{Combarro}\inst{5} \and
         \firstname{Ross} \lastname{Duncan}\inst{3,6,7}     
        % etc.
}

\institute{OpenLab, CERN, Geneva, Switzerland
\and
           Department of Physics, EPFL, Lausanne, Switzerland
\and
           Cambridge Quantum Computing Ltd, Cambridge, UK
\and 
            Department of Computer Science and Technology, University of Cambridge, Cambridge, UK
\and
            Department of Computer Science, University of Oviedo, Oviedo, Spain
\and        
            Department of Computer and Information Sciences, University of Strathclyde, Glasgow, UK
\and       
            Department of Physics, University College London, London, UK
}

\abstract{%
  Generative models, and Generative Adversarial Networks (GAN) in
  particular, are being studied as possible alternatives to Monte
  Carlo simulations.  It has been proposed that, in
  certain circumstances, simulation using GANs can be sped-up
  by using quantum GANs (qGANs).
    
  We present a new design of qGAN, the dual-Parameterized Quantum
  Circuit (PQC) GAN, which consists of a classical discriminator and two
  quantum generators which take the form of PQCs. The first PQC learns
  a probability distribution over $N$-pixel images, while
  the second generates normalized pixel intensities of an individual
  image for each PQC input.

  With a view to HEP applications, we evaluated the dual-PQC
  architecture on the task of imitating calorimeter outputs,
  translated into pixelated images.  The results demonstrate that the
  model can reproduce a fixed number of images with a reduced size as
  well as their probability distribution and we anticipate it should
  allow us to scale up to real calorimeter outputs.  }
\maketitle
\section{Introduction}
\label{intro}

The next High Luminosity Large Hadron Collider (HL-LHC) phase will
collect an overwhelming amount of data, with complex physics and small
statistical error.  To analyse this data, high precision methods which
use only limited resources are needed.  Traditional Monte Carlo based
simulation, such as \textit{Geant4} \cite{Geant4, Geant4_recent} and
the \textit{GeantV} prototype \cite{GeantV} for \textit{full
  simulation} of particle transport, is however very time-consuming,
therefore new approaches using deep neural networks have been studied
for \textit{fast simulations}.

Generative Adversarial Networks (GAN) are a strong candidate for such fast simulations. Based on two neural networks, generator and discriminator, trained alternatively, GANs have been widely explored thanks to their ability to generate images with complex structures at much high speed.  In HEP, the variations of GAN, such as CaloGAN \cite{CaloGAN} and 3DGAN \cite{3DGAN}, have achieved similar performance as full Monte Carlo based simulation, but with reduced time taken. 
 
At the same time, quantum computing has emerged as another important pillar in modern research attracting the attention of many researchers due to its potential to execute certain tasks with an exponentially reduced amount of resources both in time and space compared to classical processors \cite{Google}. It has already shown promising results in various fields, such as optimization \cite{quantum_optimization, QUBO_grover} and  cryptography \cite{cryptography}.

Advances in both deep learning and quantum computing suggest to merge them to benefit their advantages at once, leading to a new field of study, so-called \textit{Quantum Machine Learning} (QML). Quantum Generative Adversarial Networks, which are the quantum version of GANs, are one of its examples. Several quantum GAN models have been investigated in the last few years, but the scientific community still confronts a need to further explore in order to apply the model to more realistic use-cases. 

In this paper, we propose for the first time a \emph{Dual Parameterized Quantum Circuit GAN} model (dual-PQC GAN model) as one of the improvements to overcome the remaining limitations of quantum GANs. This model uses two parametrized quantum circuits, which share the role of a single quantum generator: the first PQC learns the distribution over image samples, while the second PQC determines the amplitude distribution over pixels on a single image. Thanks to this separation, it is possible to exploit the continuous nature of probability distributions over output states in quantum circuits to represent continuous variables. 

This paper is organized as follows. \secref{sec:GAN_in_HEP} summarizes the application of Generative Adversarial Networks in HEP. We then present a short overview of  a quantum version of GAN in \secref{sec:QGAN}. In \secref{sec:dual_PQC} and \secref{sec:dual_PQC_training}, the first prototype of a dual-PQC GAN model is proposed, with the results of its simulation. This paper concludes with \secref{sec:conclusion}, which summarizes and gives an outlook of future works.  

\section{Applications of GANs in HEP} 
\label{sec:GAN_in_HEP}
GANs, designed by I.\ Goodfellow et al.\ in 2014 \cite{GAN}, are
deep generative models which aim to reproduce new data from a given
original training set. They are characterized by two deep neural
networks, \textbf{Generator} $G$ and \textbf{Discriminator} $D$, which
are trained alternatively. During the training, $G$ progressively
generates data similar to the real one, while $D$ increases the
probability of assigning the correct labels to both real and fake
data. Numerous improvements have been made since the initial proposal
and, in particular, GANs have achieved remarkable success in image
processing and generation, via variations such as Deep Convolutional
GAN (DCGAN) \cite{DCGAN}, Auxiliary Classifier GAN (ACGAN)
\cite{ACGAN}, Progressive GAN \cite{PGAN}, etc.

% This is equivalent to playing a min-max optimization process between $G$ and $D$ with value function $V(G, D)$ \cite{GAN} : 
% \begin{equation}
% \min_G \max_D V(G,D) = \min_G \max_D  \big[\underbrace{\mathbb{E}_{\mathbf{x}\sim p_g(x)}[\log D_\theta(\mathbf{x}) ]}_{\text{likelihood to assign}\atop\text{correct label to real samples}} + \underbrace{\mathbb{E}_{\mathbf{z}\sim p_z}[\log(1 -D_\theta(G_\phi(\mathbf{z})))]}_{\text{likelihood to assign}\atop\text{correct label to generated samples}}\big],
% \label{eq:V_function}
% \end{equation}
% where the terms on the right hand side represent the likelihood that the discriminator assigns correct labels to real and generated data, respectively. At the end of the training, $D$ and $G$ reach the Nash equilibrium where the fake data is completely indistinguishable from the real one.

The evolution of GANs has attracted strong interest in the high energy physics domain. In HEP, the detectors can be described as 3D cameras, recording pictures of particle collisions. Calorimeters, in particular, measure the energies deposited by a shower of the particles that traverse them. They generally consist of alternate arrays of active sensor material and passive dense layers to ensure that the incoming (primary) particle will deposit most of its energy inside their volume. These energy depositions can be compared to the monochromatic pixel intensities of a 3D image. Because of their high granularity, the detailed simulation of a calorimeter is particularly time-consuming. As a result, GANs come into the limelight to allow fast simulation of particle showers with high fidelity.       
% The primary particle creates a shower of secondary particles as it passes through the detector.
% Location Aware GAN (LAGAN) \cite{LAGAN} and CaloGAN \ cite{CaloGAN} are among the first applications of DCGAN in HEP, designed for simulating high granularity calorimeters. LAGAN is specialized in reconstructing sparse and highly non-linearly location-dependent two-dimensional calorimeter data via locally connected layers. Furthermore, following the similar strategy as ACGAN, its discriminator has an additional classification task calculating the total energy deposit in the pixels.  CaloGAN, which uses multiple  LAGAN, reconstructs 3D energy shower in a multi-layer electromagnetic calorimeter by concatenating 2D image layers, each of which generated by one LAGAN.  

One possible application of GAN in HEP is 3DGAN \cite{GAN_fast_simulation, 3DGAN}, which is a 3D extension of GAN, using 3D (de-)convolutional layers to capture the whole 3D energy profile. It simultaneously performs two additional tasks of estimating the incoming particle energy and measuring the total deposited energy to enhance the stability and convergence of networks. Details on 3DGAN architecture and its performance validation  are available in  \cite{3DGAN}. 

% \begin{figure}[h]
%   \centering
%   \begin{subfigure}{0.4\textwidth} 
%     \includegraphics[width = \textwidth]{CaloGAN_real.eps}
%     \caption{\em Real image.}
%   \end{subfigure}
%   \begin{subfigure}{0.4\textwidth} 
%     \includegraphics[width = \textwidth]{CaloGAN_gen.eps}
%     \caption{\em Generated image.}
%   \end{subfigure}
%   \caption{\em $x-z$ projection of 3D energy shower (averaged over 10,000 samples) in electromagnetic calorimeters reproduced by Geant4 (a) and 2D version of 3DGAN (b).  }
%   \label{fig:3DGAN_image}
% \end{figure}

% To fully validate results of 3DGAN, \figref{fig:3DGAN_image} compares 2D projection of 3D particle shower profiles in a  electromagnetic calorimeter prototype \cite{CLIC} generated by (a) traditional \textit{geant4} and (b) by the 2D version of 3DGAN. The image profiles produced by \textit{geant4} are also used as GAN training set. The results show that GAN can reproduce the jet images faithfully, justifying its usage in HEP for fast simulations. 

\section{Quantum Generative Adversarial Networks} 
\label{sec:QGAN}

% Quantum machine learning (QML) algorithms, including QGAN, are among the algorithms which can be efficiently performed on Noisy Intermediate-Scale Quantum (NISQ) real hardware.  
The possibility of combining machine learning and quantum computing also led to the generalization of GAN to quantum systems by S. Lloyd and C. Weedbrook \cite{QGAN}. The main mechanism of the model, the adversarial training,  is reproduced, but different scenarios are possible:  the input data can be either quantum data or classical data embedded in quantum states and the discriminator/the generator can also be either classical or quantum. 

% Suppose that the real data can be expressed by a set of quantum states with density matrix $\sigma$. Then, the generator, constructed in the quantum system, aims to reproduce a density matrix $\rho$, which is as similar as possible to $\sigma$. 

Since its initial proposal, several quantum GAN (qGAN) variations have
been suggested to generate either classical data \cite{QGAN_qiskit,
  QGAN_Romero, QGAN_Situ} or quantum data \cite{QGAN_Pierre, Hu_2019,
  Benedetti_2019, OpticalGAN}. 
Zoufal et al~\cite{QGAN_qiskit} proposed a hybrid qGAN model composed
of a quantum generator and a classical discriminator to train on
\emph{classical} data. During the training, the generator learns an
arbitrary probability distribution over discrete variables, which is
encoded in the amplitudes of the final quantum state. This model was
applied of quantum finance, and demonstrated using the real quantum
hardware, \textit{IBM Q Boeblingen}. Anand et al~\cite{QGAN_experimental} also present similar results, but simulated on the real Rigetti quantum hardware, \textit{Aspen-4-2Q-A}.  
Unlike the aforementioned models treating classical data, Situ et al \cite{QGAN_Situ} propose a qGAN model which aims to approximate an unknown pure quantum state with a quantum generator and a quantum discriminator. One problem with quantum machine learning models is the apparent difficulty of training PQCs, captured by the vanishing gradient and barren plateau problems. Fortunately, there have been also studies on the methods to improve the performance of qGAN, for instance, Quantum Multiplicative Matrix Weight (QMMW), which helps to avoid mode collapse or vanishing gradient problem in qGAN \cite{QMMW}.   

The results from the previous research are impressive, showing the potential of qGAN for the near-term quantum hardware. However, additional investigations are needed in order to fully understand the quantum advantages of qGAN and generate not only simple probability distributions but also  more complex image samples. 

\section{Dual-PQC GAN model} 
\label{sec:dual_PQC}

\begin{figure}[b]
  \centering
  \begin{quantikz}
    \lstick{} &  \gate{R_y(\phi^0_0)}\gategroup[3,steps=1,style={dashed,
      rounded corners,fill=yellow!20, inner xsep=1pt, inner ysep = 1pt},background, label style={label position=below,anchor=north,yshift=-0.3cm} ]{{\sc Initialization}} & \qw & \ctrl{1}\gategroup[3,steps=3,style={dashed,
      rounded corners,fill=blue!20, inner xsep=1pt, inner ysep = 1pt},background, label style={label position=below,anchor=north,yshift=-0.3cm} ]{{\sc layer $1$}} &\qw & \gate{R_y(\phi^0_1)} & \qw & \ldots &  & \ctrl{1}\gategroup[3,steps=3,style={dashed,
      rounded corners,fill=blue!20, inner xsep=1pt, inner ysep = 1pt},background, label style={label position=below,anchor=north,yshift=-0.3cm} ]{{\sc layer $k$}} &\qw & \gate{R_y(\phi^0_k)} &\qw \\ [-0.4cm]
    \lstick{} &  \gate{R_y(\phi^1_0)} & \qw & \control{} & \ctrl{1} & \gate{R_y(\phi^1_1)}  &\qw & \ldots &  & \control{} & \ctrl{1}  & \gate{R_y(\phi^1_k)} &\qw\\[-0.4cm]
    \lstick{}&  \gate{R_y(\phi^2_0)} & \qw & \qw & \control{} & \gate{R_y(\phi^2_1)} & \qw & \ldots & & \qw & \control{} & \gate{R_y(\phi^2_k)} &\qw 
  \end{quantikz}
  \caption{\it  Quantum variational form using Pauli-$Y$ rotations and $CZ$ gates with depth $k$.}
  \label{fig:vf}
\end{figure}
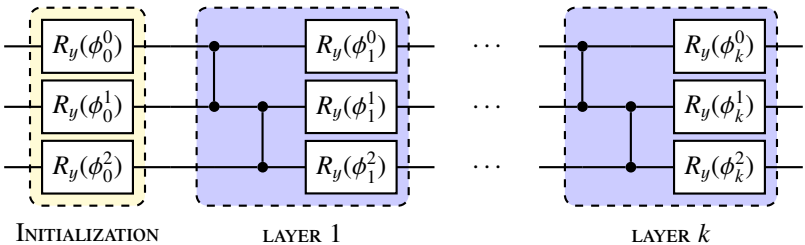

Our  preliminary experiments involved training a qGAN, as conceptualised in \cite{QGAN_qiskit}, for the calorimeter problem -- however this immediately revealed a problem: as the image itself is encoded in the amplitudes of the computational basis states, this meant that only the average of all the training samples could be learned, and the GAN did not, therefore, sample typical images. To put this in more precise terms, in order to achieve the exponential compression of representing a $N = 2^n$ pixel image using $n$ qubits, it follows that the qGAN prepares a quantum state of the form:
\begin{equation}
\ket{\psi} =  {\textstyle\sum}_{j=0}^{2^{n}-1}\sqrt{I_{j}}\ket{j}
\end{equation}
where $I_j$ is the intensity of the $j^{th}$ pixel. So we can see
that, in a sense, the qGAN encodes a \textit{single} image as
a probability distribution, and so there is no room left to also
encode a probability distribution, representing the full dataset, to
sample images from. This problem does not arise in classical GANs,
where a single sample from a $\mathcal{O}(N)$ sized neural network
generates a single image in one go. 

In this section we describe our solution, a new type of quantum GAN,
the \textit{dual Parameterized Quantum Circuit (PQC) GAN model}
(dual-PQC GAN model). It aims to reproduce a set of image samples from
real training data while preserving the exponential compression
achieved by amplitude encoding.  The work in
\cite{rudolph2020generation} follows from similar motivation.

The dual-PQC GAN is a hybrid qGAN architecture which has one classical
discriminator and two parameterized quantum circuits, PQC1 and PQC2,
sharing the role of the generator.  PQC1, with $n_1$ qubits, learns
the probability distribution over image samples and PQC2, with $n_2$
qubits, learns the amplitude distribution over pixels of each
image.  The classical discriminator takes the training set and the
images generated by PQC2, and it classifies them into real and
fake. The predicted labels are used to tune alternatively $\phi_1$,
$\phi_2$ and $\theta$, the parameters for PQC1, PQC2, and the
discriminator, respectively.

In this study, both PQCs consist of alternating layers of
single-qubit Pauli-rotations and a set of two-qubit entanglement
gates, as shown in \figref{fig:vf} It is widely used in quantum
machine learning thanks to its strong expressive power, offering an
effective way of reconstructing an expected behaviour
\cite{variational_form, PQC}. We use RY rotation gates and CZ
entanglement gates, but other choices are possible.

 \tikzstyle{Generator2} = [rectangle, draw, fill=green!20, 
text width=4.6em, text centered, rounded corners, minimum height=1.3cm]
\tikzstyle{Discriminator2} = [rectangle, draw, fill=blue!20, 
text width=6.8em, text centered, rounded corners, minimum height=1cm]
\tikzstyle{input2} = [rectangle, draw, fill= white, 
text width=0.3em, minimum height = 2.3em, align= center,
copy shadow={draw, shadow xshift=1mm, shadow yshift=-1mm} ]
\tikzset{meter/.append style={draw, inner sep=4, rectangle, font=\vphantom{A}, minimum width=16, line width=.36,
    path picture={\draw[black] ([shift={(.04,.12)}]path picture bounding box.south west) to[bend left=50] ([shift={(-.04,.12)}]path picture bounding box.south east);\draw[black,-latex] ([shift={(0,.04)}]path picture bounding box.south) -- ([shift={(.12,-.04)}]path picture bounding box.north);}}}
    
\tikzstyle{line} = [draw, -latex']
\begin{figure}[h]
  \begin{center}
    \begin{tikzpicture}[align=center,node distance = 2cm, auto]
    % Place nodes

    \draw (6.5,-0.5) node[input2, label = below:{
      Fake Data \\ by sampling}] (FakeData) {\phantom{input}};
    \node [Discriminator2, right of = FakeData, node distance=2.3cm] (Disc) {\bf   Classical \\ Discriminator}; 
    \node [input2, above of = Disc, node distance=1.3cm, label = right:{Real Data}] (RealData) {\phantom{input}};
    \node [text width = 3.5em, right of = Disc, node distance = 2.5cm, minimum height = 1.6cm](PredictedLabel) {Predicted Labels};
    \node [text width = 3.8em](Discard) at (7,0.5) {Discard};
    \node [rectangle, text width=6em, minimum height = 1em, align= center] (Basis) at (0,-0.5) {${\ket{0},...,\ket{2^{n_1}-1}}$};
    \node [rectangle, text width=4em, minimum height = 1em, align= center] (Zeros) at (0,0.5) {$\ket{0}^{\otimes n_2 - n_1}$};
    \node [Generator2, below of = Basis, node distance = 1.9cm] (PQC1) { { \textbf{PQC1}} \\  \footnotesize $n_1$ qubits \\ depth $d_{g,1}$};
    \node [Generator2] (PQC2) at (3,0) {{ \textbf{PQC2}} \\ \footnotesize $n_2$ qubits \\ depth $d_{g,2}$};
    
    \draw[thick] (4.4, 0.35) -- (4.7, 0.65);
    \draw[thick] (4.4, -0.65) -- (4.7, -0.35);
    
    \draw[thick] (1.4, 0.35) -- (1.7, 0.65);
    \draw[thick] (1.4, -0.65) -- (1.7, -0.35);
    
    \node[text width = 5em] at (4.55,0.72) {$n_2 - n$};
    \node[text width = 1em] at (4.55,-0.28) {$n$};
    
    \node[text width = 5em] at (1.55,0.72) {$n_2 - n_1$};
    \node[text width = 1em] at (1.55,-0.28) {$n_1$};
    % Draw edges
    \path [line] (PQC1) -- (Basis);
        \draw (Basis)  |- (PQC2.west |-Basis.east); 
        \draw (Zeros)  |- (PQC2.west |-Zeros.east); 
    \path [line] (PQC2.east|-FakeData.west) |- (FakeData);
    \path [line] (PQC2.east|-Discard.west) |- (Discard);
    \path [line] (FakeData.east) -- (Disc);
    \path [line] (Disc) -- (PredictedLabel);
    \path [line] (RealData) -- (Disc);
    \path [line, style = dashed] (PredictedLabel.south) -- ++(0,-1.1cm)  -| (Disc.south);
    \path [line, style = dashed] (PredictedLabel.south) -- ++(0,-1.1cm)  -|node [pos = 0.25, below] (TextNode) {Classical Optimization}  (PQC2.south);
    \path [line, style = dashed] (PredictedLabel.south) -- ++(0,-1.1cm)  -- (PQC1.east);
    
    \node[meter, above of = PQC1, node distance = 1.1cm](meter){};
    \node[meter, left of = Discard, node distance = 1.45cm](meter){}; 
    \node[meter, left of = FakeData, node distance = 0.95cm](meter){};      
    \end{tikzpicture}
  \end{center}
  \caption{\em Schematic Diagram of dual-PQC GAN to reproduce images of $2^n$ pixels. 
  % PQC1 learns the distribution over image samples and passes a computational basis to PQC2 to "seed" an image. PQC2, taking a computational basis as input, measures $n$ qubits among its $n_2$ qubits, and returns a normalized image for each input by constructing the probability distribution over $2^n$ states, each of which corresponds to one pixel in an image. In order to avoid unitarity problem, the number of qubits in PQC2, $n_2$, should be greater than $n$, and the remaining $n_2 - n$ qubits are discarded. The images reproduced by PQC2, following the distribution obtained from PQC1, are passed to a classical discriminator. The predicted labels are then used to tune parameters for PQC1, PQC2, and classical discriminator. The model can eventually reproduce $2^n$ images of $2^n$ pixels.
  }
  \label{fig2}
\end{figure}
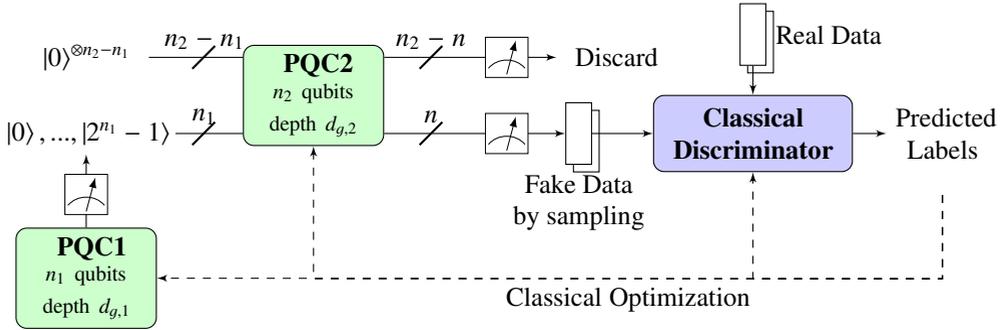
Consider a training set $X \subset \mathbb{R}^{2^n}$ of $N = 2^n$
pixel images. To begin, the output state of PQC1 is measured
producing $n_1$ bits.  Then, via a set of Pauli-$X$ gates, this bit string
is used to initialise PQC2 with the corresponding computational basis
state in the $2^{n_1}$ dimensional Hilbert space, $\ket{i} \in
\{\ket{0},..., \ket{2^{n_1} - 1}\}$.  Then, by repeatedly measuring
$n$ output qubits of PQC2, the probability distribution over the
computational basis in the $2^{n}$ dimensional Hilbert space, $\ket{i}
\in \{\ket{0},..., \ket{2^n - 1}\}$, is constructed and translated as
an image of size $2^n$ for each input state. 

Note that since PQC2 performs a unitary operation, and since its
inputs are always computational basis states, the output quantum
states are necessarily orthogonal. This puts an unwanted restriction
on the possible images.  Relying on the Stinespring dilation theorem,
we can remove this restriction by choosing $n_2 > n_1, n$ and using
some ancilla qubits which are discarded at the end.  We will return to
this point at the  end of the  section.

%Let $X = \{x | x \in [0,... N - 1]\}$  the training set of real values between $0$ and $N-1$, following a distribution $p_{real}$.  The $n$-qubit quantum generator $G_\phi$ transforms the input state $\ket{\psi_{in}}$ in Hilbert space of dimension $N = 2^n$ into the final state $\ket{g(\phi)}$ : 
% \begin{equation}
% G_\phi\ket{\psi_{in}} = \ket{g(\phi)} = \sum_{i = 0}^{N  - 1}\sqrt{p_g^i(\phi)}\ket{i}
% \end{equation}
% where $p_g^i(\phi)$ denotes the probability of state $\ket{i}$ in the final state and $\{\ket{i}\}_{i = 0,...,N-1}$ the computational basis in $N$-dimensional Hilbert space. 

Let $p_g^i$ be the probability that state $\ket{i}$ is
measured by PQC1; let $\mathcal{I}_i$ denote the normalized image
produced by PQC2 when given the input state $\ket{i}$; and  let $I_{ij}$ be
the amplitude of $j^{th}$ pixel of $\mathcal{I}_i$ with $i\in \{0,..., 2^{n_1}-1\}$ and $j \in \{0,..., 2^n-1\}$. Then the output states generated by PQC1, $G_{1,\phi_1}$, and PQC2, $G_{2,\phi_2}$, are explicitly given as : 
\begin{equation}
G_{1,\phi_1}\ket{\psi_{initial}} = \ket{g_{1,\phi_1}} = \sum_{j=0}^{2^{n_1}-1}\sqrt{p_g^j}\ket{j},
\end{equation}
\begin{equation}
%G_{2,\phi_2}\ket{i} = \ket{g_{2,\phi_2}} = %\sum_{j=0}^{2^{n}-1}\sqrt{I_{ij}}\ket{j}\otimes\sum_{j=0}^{2^{n-n_2}-1}c_j\ket{j},
%SJH alternative -- original still kept in comments above.
G_{2,\phi_2}\ket{\mathbf0}\ket{i} = \ket{g_{2,\phi_2}} = \sum_{j=0}^{2^{n}-1}\sqrt{I_{ij}}\ket{\Psi_j} \ket{j}
\end{equation}
where $\ket{\psi_{initial}}$ is the input state of PQC1 fixed during the whole training and $\ket{\Psi_j}$ are some $n-n_2$ qubit states that we discard.%$c_j \in [0,1]$. 

During the training, PQC1 learns the distribution $p_g(i)$ over $\mathcal{I}_i$, so that it approaches to the real distribution, $p_{real}$ over $X$. On the other hand, PQC2 learns the  amplitude over $2^n$ pixels for $2^{n_1}$ images, to make $\mathcal{I}_i$ as close as possible to real images. At the end of the training, the true/fake probabilty  predicted by the discriminator for $\mathcal{I}_i$, $D(\mathcal{I}_i)$ should converge to $1/2$.

For the following simulations, we use a modified min-max loss, given as:  

\begin{equation}
L_G =  -\frac{1}{m}\sum_{i = 1}^m \log D(G( \mathbf{z}_i)) = -\sum_{i=0}^{2^{n_1} - 1} p_g^i\log(D(\mathcal{I}_i))
\label{eq:lg}
\end{equation}
\begin{equation}
L_D = \frac{1}{m} \sum_{i = 1}^{m} \big[\log D(\mathbf{x}_i) + \log ( 1 - D(G( \mathbf{z}_i)))\big] = \frac{1}{m} \sum_{i = 1}^{m} \log D(\mathbf{x}_i) + \sum_{i=0}^{2^{n_1} - 1} p_g^i\ \log ( 1 - D(\mathcal{I}_i)) ,
\label{eq:ld}
\end{equation} 
where $m$ is the batch size, $x_i$ the real data and $z_i$ random input. The first equality gives the definition of the loss in the classical GAN and the second equality the practical formula used in dual-PQC GAN simulations. 

% In practice, the usual binary cross entropy loss for PQC1 and PQC2 can be also rewritten as: 
% \begin{equation}
%   L_G = -\sum_{i=0}^{2^{n_1} - 1} p_g(i)\log(D(\mathcal{I}_i)).
%   \label{eq:lg} 
% \end{equation}

Based on the calculated loss, the parameters in the quantum generator are tuned by computing the analytic quantum gradient descent \cite{QGD}, while the discriminator is optimized in the exactly same way as in classical GAN. Further details on the method used for analytic quantum gradient are explained in Ref. \cite{QGD2, QGD3}. 
%{\color{red}The total complexity of dual PQC GAN model is . }

Ultimately, the dual-PQC GAN model can generate $2^{n_1}$ images of
size $2^n$. Increasing the number of qubits used in PQC1 and PQC2
allows to increase both the number and size of produced images. This
model shows an advantage in terms of computational resources
by using only $\mathcal{O}(\log(N))$ qubits,     compared to the
classical neural networks with $\mathcal{O}(N)$ neurons to reproduce
an image of size $N$. Specifically, the potential advantages are
threefold: firstly, there is an exponential reduction in space
(memory) requirement; secondly, the resultant exponential reduction in
number of tunable parameters (i.e. the number of tunable parameters is
proportional to the number of gates in the PQC or neurons in the
classical neural network) suggests that training could be performed
more efficiently; finally, it is potentially advantageous to have the
images encoded in quantum states if further processing is to be
performed (for example if that image processing can itself be
performed more efficiently using quantum computing). It should,
however, be noted that if all one wants to do is to generate
an image, then the number of samples required from PQC2 is exponential
in the number of qubits.

\paragraph{How many ancillas are needed?} 

Since we want to be able to reproduce any collection of images, we
have to use some ancillary qubits.  Stinespring's theorem gives an
upper bound on the number of ancillas but it is not tight.  We will
now show that $n_2 = 2n$ will suffice when we assume for simplicity that $n_1 = n$. 

Let $U$ denote the unitary
matrix corresponding to PQC2.  When $U$ is applied to the input state
$\ket0^{\otimes n}\otimes \ket i$, where $\ket i$ is an $n$-qubit basis
state, the resulting state is always one of the columns of $U$; in
particular it always one of the first $2^n$ columns.  We'll construct
an example of $U$ which realises $2^n$ arbitrary images.  Let 
\[
\ket{\mathcal{I}_i}
= (\sqrt{I_{i,0}}e^{i\phi_{i,0}}, \ldots, \sqrt{I_{i,0}}e^{i\phi_{i,0}})^T
\]
be a state whose amplitudes encode the pixels of image $i$.  Consider
the state $\ket{U(i)} = \ket i \otimes \ket{\mathcal{I}_i}$, and
observe that if its first $n$ qubits are measured in the computational
basis and discarded, then the remaining quantum state is precisely
$\ket{\mathcal{I}_i}$.  Since $\bra{U(i)}\ket{U(j)} = \delta_{ij}$, ie, they are orthonormal, we
can construct the required \textit{unitary} matrix, $U$, by setting the first $2^n$
columns to be $\ket{U(i)}$ for $i\in\{0,\ldots 2^n-1\}$ and choosing
the remainder arbitrarily.  Since these columns correspond to states
which will not be selected by any input to PQC2, they don't matter.
It should be noted that it is unlikely that the trained dual-PQC GAN
would actually converge on unitaries of such a form, and thus this
construction is given merely to demonstrate that $2n$ qubits suffice;
further improvements are surely possible.

%Alternatively, the number of qubits used in PQC2 can be reduced to $n_2 < 2n$, by giving random noise as input to the remaining $n_2 - n_1$ qubits on PQC2. However, for simplicity,  the following study will focus on the basic case with $n_2  = 2n_1 = 2n$. 

\section{Training dual-PQC GAN}
\label{sec:dual_PQC_training}

\begin{figure}[b]
    \centering
  \begin{subfigure}{0.45\textwidth}
    \includegraphics[width = \textwidth]{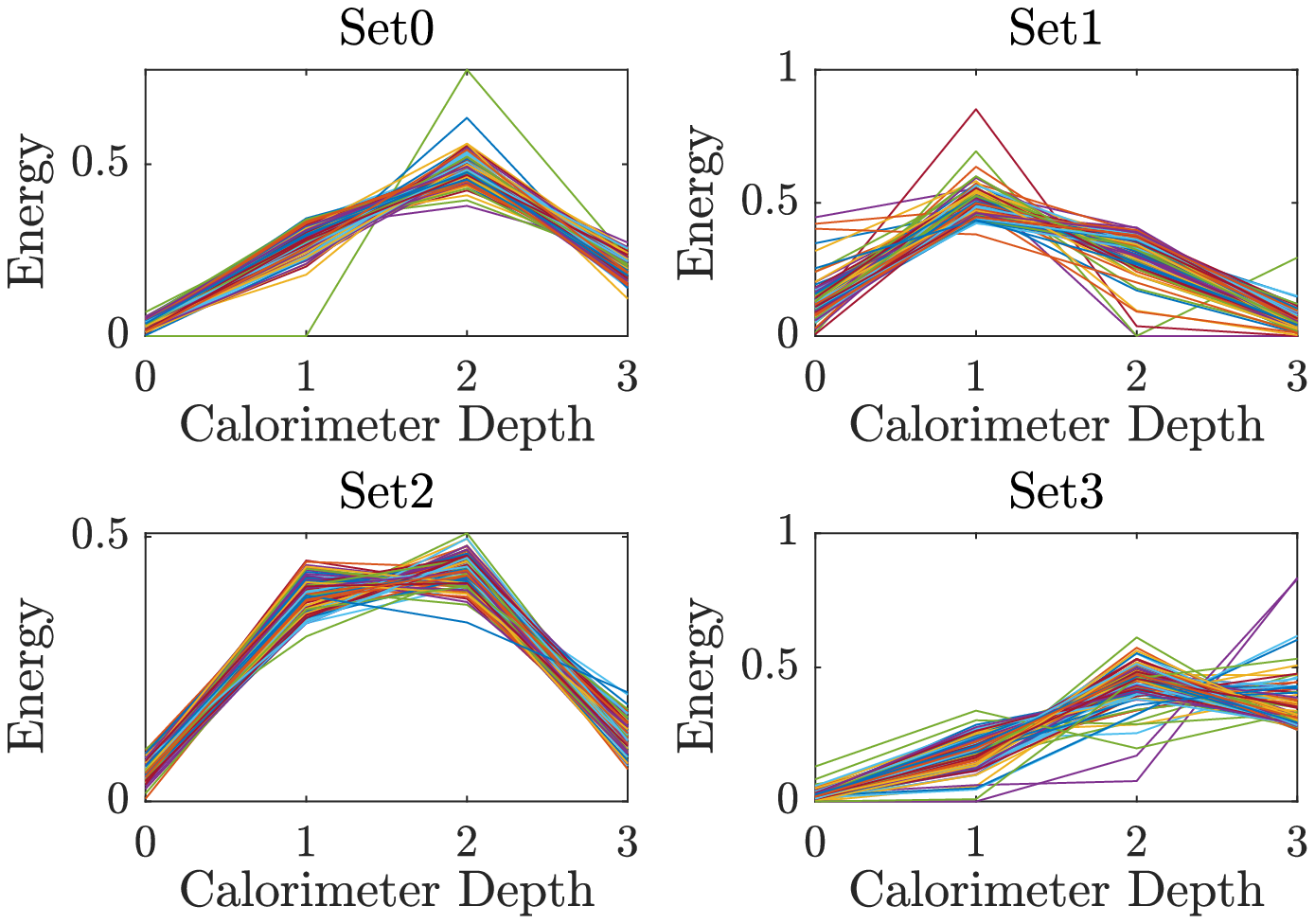}
    \caption{\em}
    \label{fig:classification}
  \end{subfigure}
  \begin{subfigure}{0.45\textwidth}
    \centering
    \includegraphics[width = \textwidth]{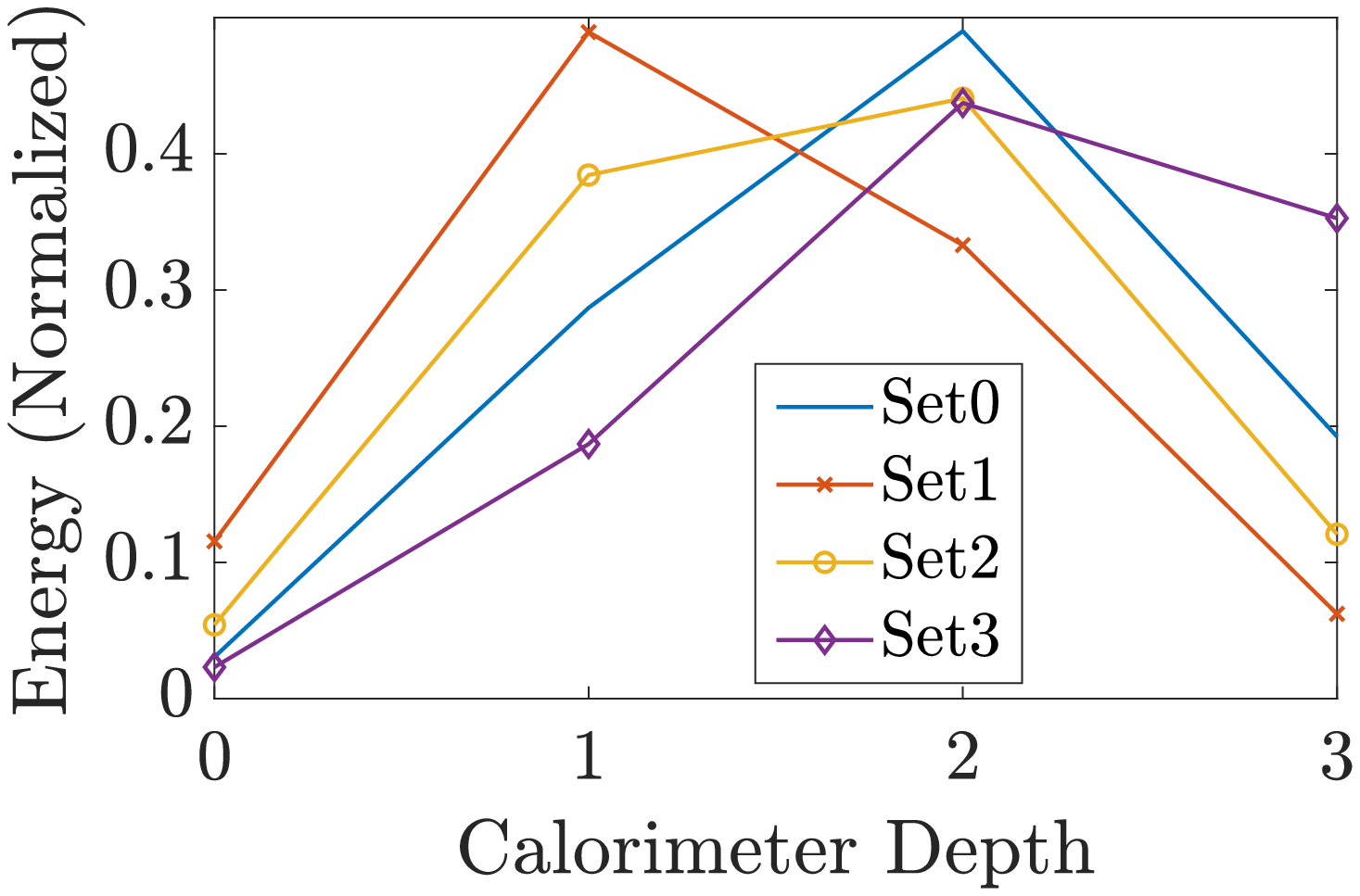}
    \caption{\em }
    \label{fig:mean_images}
  \end{subfigure}
  \caption{\em Classification of 20,000 normalized real image samples (only 100 samples displayed) into $4$ classes via K-means clustering \cite{Kmeans} (a) and their average (b). The mean image of Set $i$ is assigned to a computational basis state $\ket{i}$.}
  \label{fig:classiciation_mean}
\end{figure}

 This section tests  dual-PQC GAN described in \secref{sec:dual_PQC} and shows its potential to generate a set of image samples from a training set with a certain degree of fidelity.  We emphasize that, in order to work with a manageable number of qubits,  this study simplifies the original 3DGAN problem by reducing it to a 1D problem: reproducing the energy pattern along the calorimeter depth. In other words, the training dataset is composed by 1D energy profiles along the calorimeter $z$ dimension, averaged over $N=4$ pixels. It should be noted that such drastic reductions in problem size are common-place in quantum machine learning, in order to obtain useful proof-of-principle results.
 
 In order to evaluate the performance of the model, the original data set of 20,000 sample images is classified into $2^{n}$ classes of via K-means clustering \cite{Kmeans} as shown on \figref{fig:classification} for the case $n = 2$. The average image for each class, displayed on \figref{fig:mean_images}, gives an insight on the shape of images which should be produced by PQC2. Note that this clustering is purely for evaluation of results - raw data are used for the training.

Although, theoretically,  it is sufficient  to take $n_1 = n = 2$, several preliminary simulations have shown that $n_1 = 2n = 4$ gives better stability in the results. Therefore, for the following simulations, PQC1 takes $n_1 = 4$ but still builds a probability distribution over $2^2 = 4$ images, by only measuring $n = 2$ qubits among four. PQC1 is initialized with an equiprobable superposition over the computational basis, $\{\ket{0},..., \ket{2^{n_1}-1}\}$ with $n_1 = 4$. Furthermore, the initial parameters for both PQC1 and PQC2 are sampled from a uniform distribution over $[-\delta, \delta]$, with $\delta = 10^{-1}$. The discriminator is implemented in PyTorch, using an input layer with 4 nodes, two hidden layer with 256, 128 nodes, respectively, and a single node output layer. After the first two layers follows a Leaky ReLU function \cite{LeakyReLU} with $\alpha = 0.2$ and a sigmoid function \cite{sigmoid} is applied after the output layer. In addition, a gradient penalty \cite{penalty} for real images is added to help stability and convergence of the model, with the parameters $\lambda = 7$, $k = 0.01$ and $c = 1$. 
The dual-PQC GAN is trained using the AMSGRAD optimizer with initial learning rate of $10^{-4}$ for PQC1 and discriminator and $10^{-3}$ for PQC2.

\begin{figure}[h]
  \centering
  \begin{subfigure}{0.3\textwidth} 
    \includegraphics[width = \textwidth]{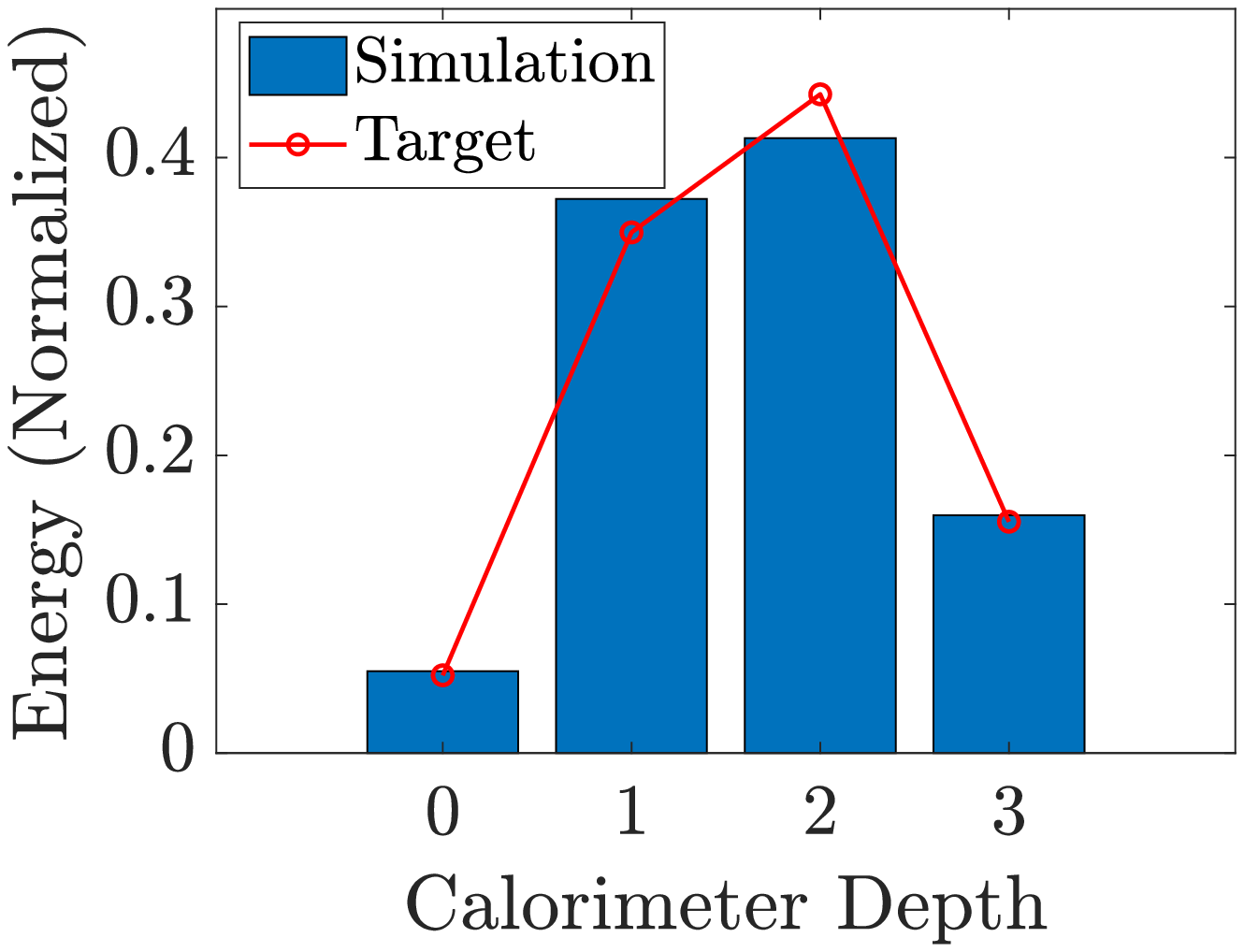}
    \caption{\em}
  \end{subfigure}
  \begin{subfigure}{0.3\textwidth} 
    \includegraphics[width = \textwidth]{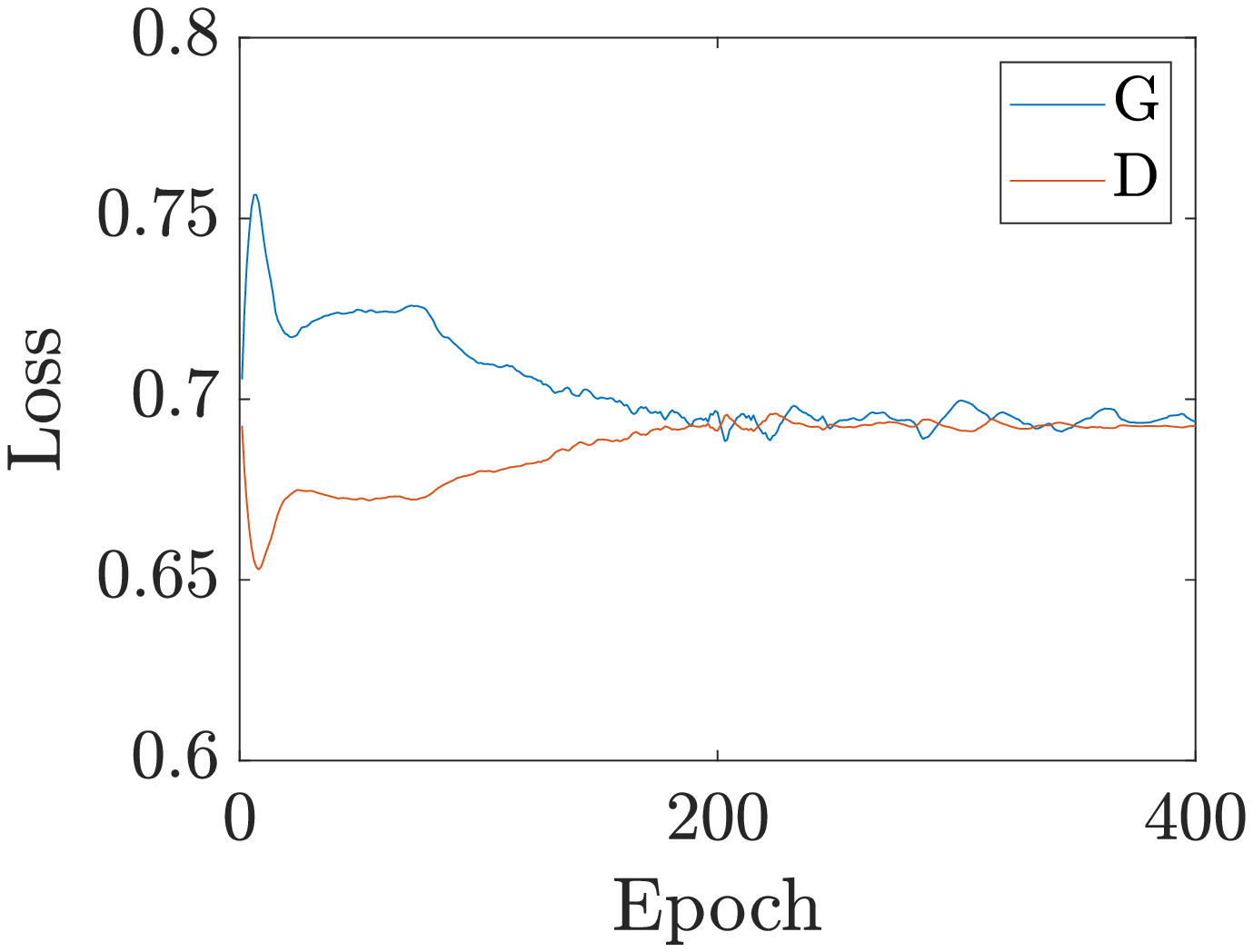}
    \caption{\em}
  \end{subfigure}
  \begin{subfigure}{0.3\textwidth}
    \includegraphics[width = \textwidth]{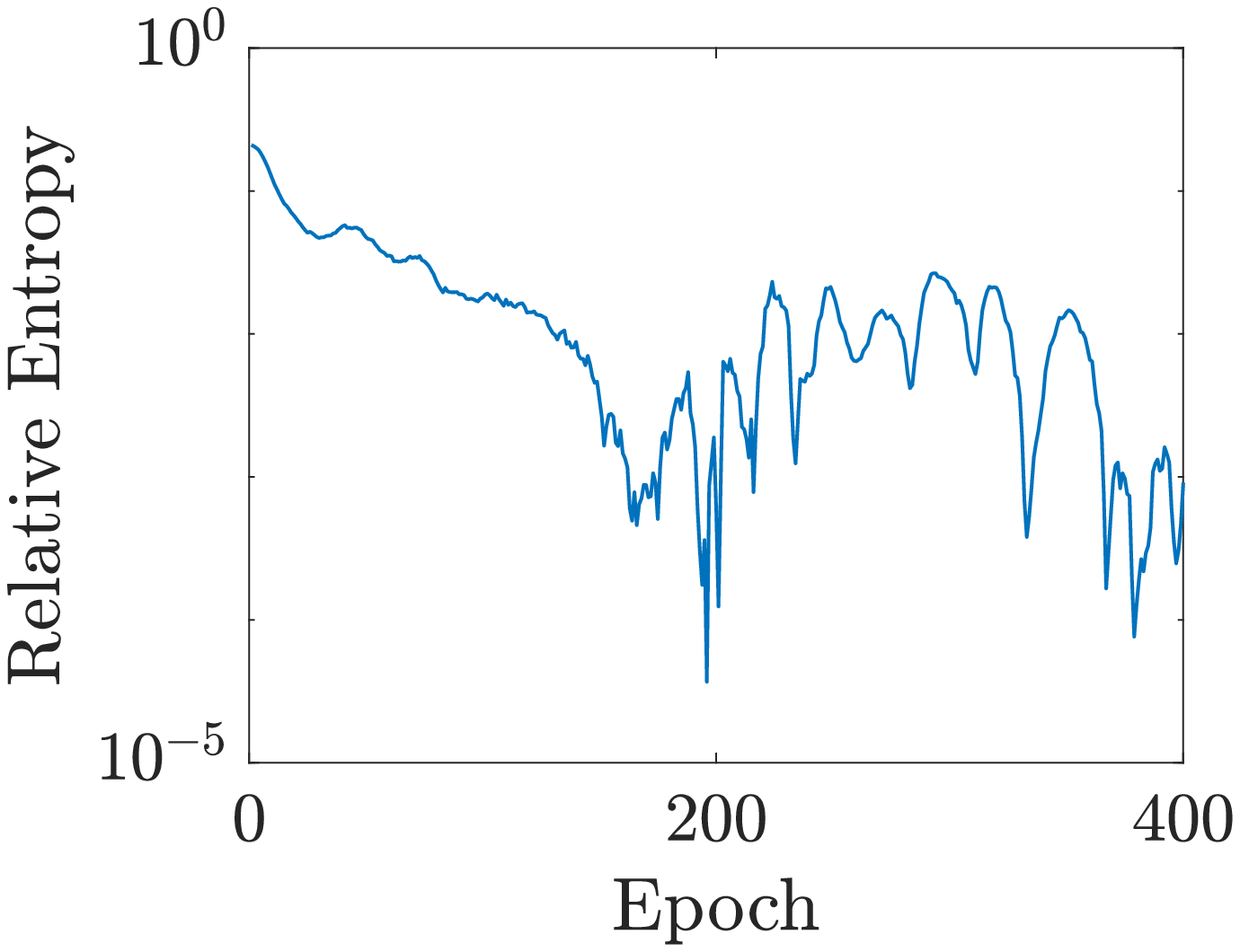}
    \caption{\em}
    \label{fig:rel_entr_N4_D2D6}
  \end{subfigure}
  \begin{subfigure}{0.3\textwidth} 
    \includegraphics[width = \textwidth]{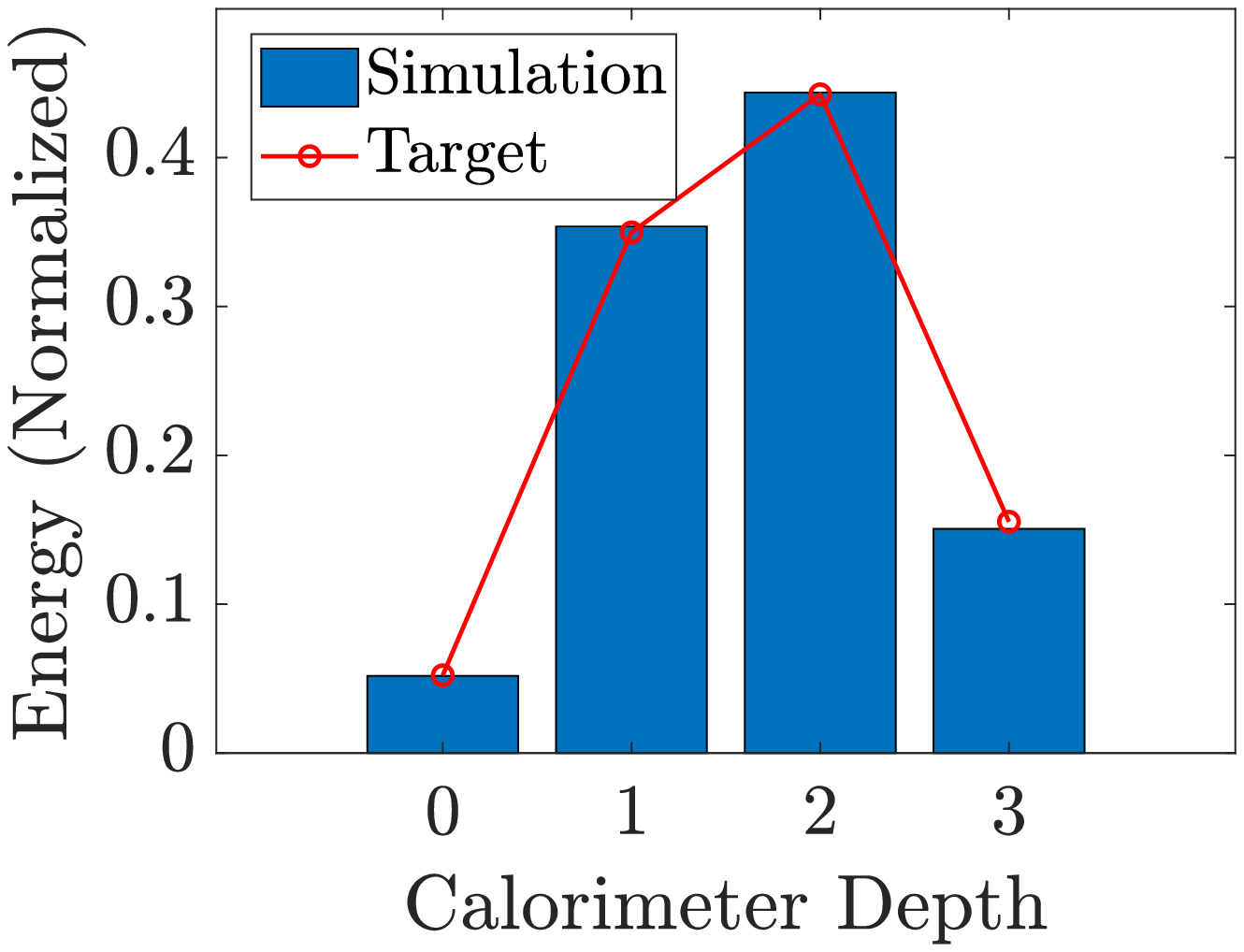}
    \caption{\em}
  \end{subfigure}
  \begin{subfigure}{0.3\textwidth} 
    \includegraphics[width = \textwidth]{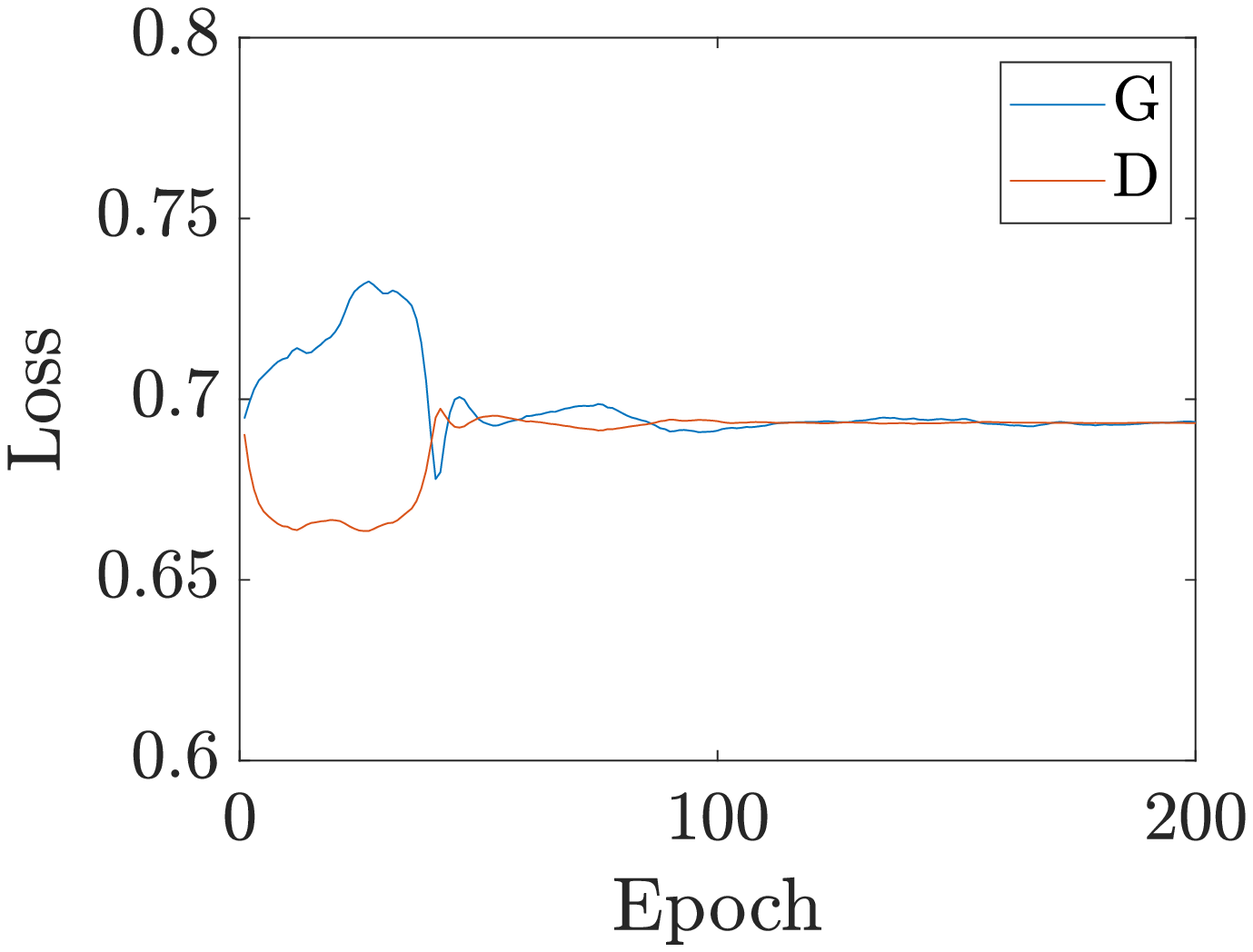}
    \caption{\em}
  \end{subfigure}
  \begin{subfigure}{0.3\textwidth}
    \includegraphics[width = \textwidth]{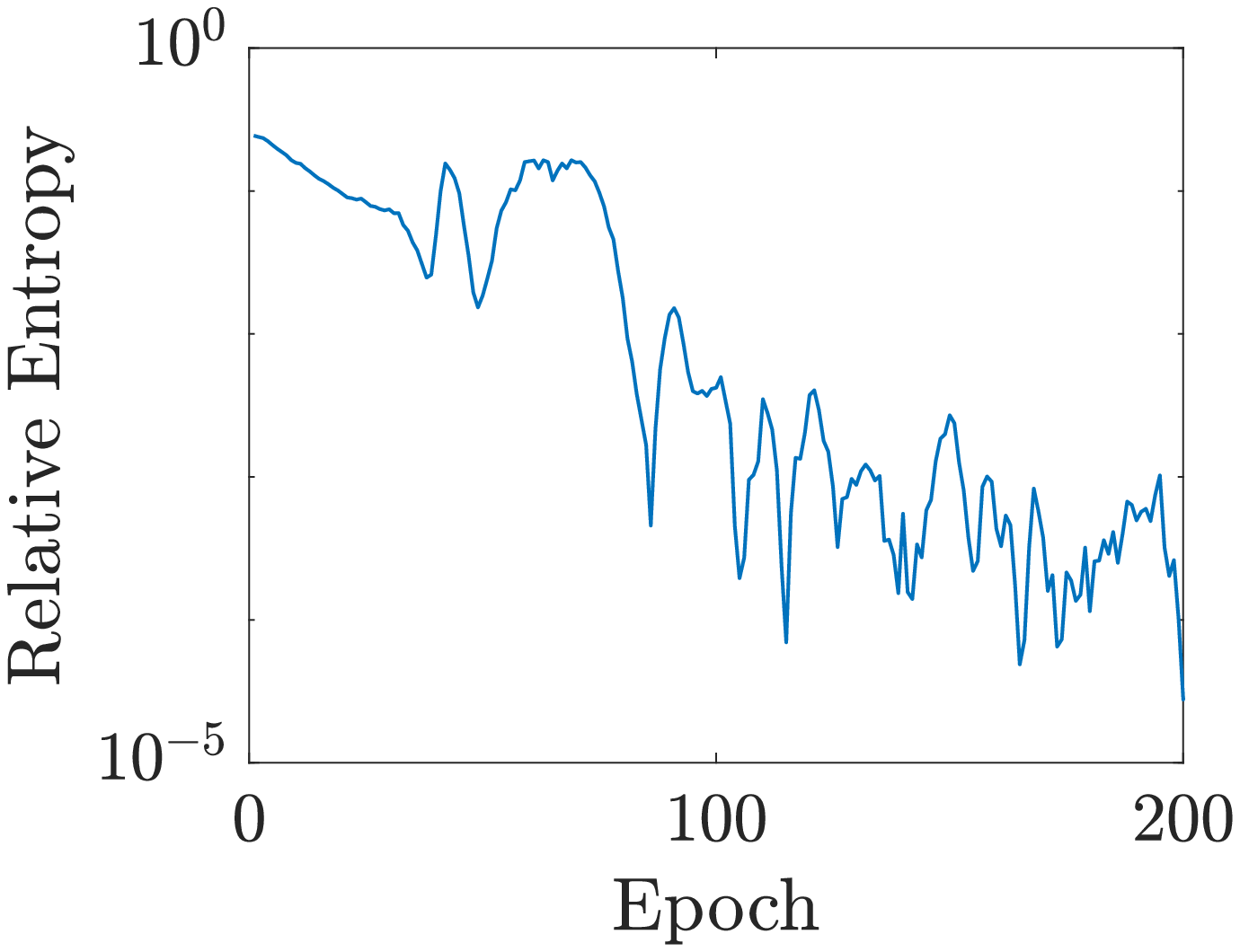}
    \caption{\em}
    \label{fig:rel_entr_N4_D2D16}
  \end{subfigure}
  
  \caption{\em Results of dual-PQC GAN simulations with $n = 2$, $n_1 = n_2 = 4$, $d_{g,1} = 2$ and $d_{g,2} = 6$ (a, b, c) / $d_{g,2} = 16$ (d, e, f). The mean image amplitudes calculated by \eqnref{eq:mean_image} (a,d), the generator and the discriminator losses (b, e), and the progress in relative entropy for the mean image (c,f) are illustrated.
  }
  \label{fig:mean_loss_dual}
\end{figure}

\figref{fig:mean_loss_dual} displays progress in the loss functions and relative entropy, as well as the average of generated images at the end of the training, weighted with their probability, given by: 
\begin{equation}
  \mathcal{I}_{mean} = {\textstyle\sum}_{i=0}^{2^n - 1}p_g^i\cdot \mathcal{I}_i.
  \label{eq:mean_image}
\end{equation}

Both cases of $d_{g,2} = 6$ and  $d_{g,2} = 16$ exhibit convergence in mean energy distribution towards the target, as well as convergence of generator and discriminator losses. Furthermore, the relative entropy between real and generated mean images reaches below $10^{-4}$ as shown on \figref{fig:rel_entr_N4_D2D6} and \figref{fig:rel_entr_N4_D2D16}. Note that after the convergence in loss function, the relative entropy does not cease decreasing in case of $d_{g,2} = 16$, while it starts to oscillate with large amplitude in case of $d_{g,2} = 6$, reflecting certain degree of instability in the simulation.

\begin{figure}[h]
  \centering
  \begin{subfigure}{0.245\textwidth} 
    \includegraphics[width = \textwidth]{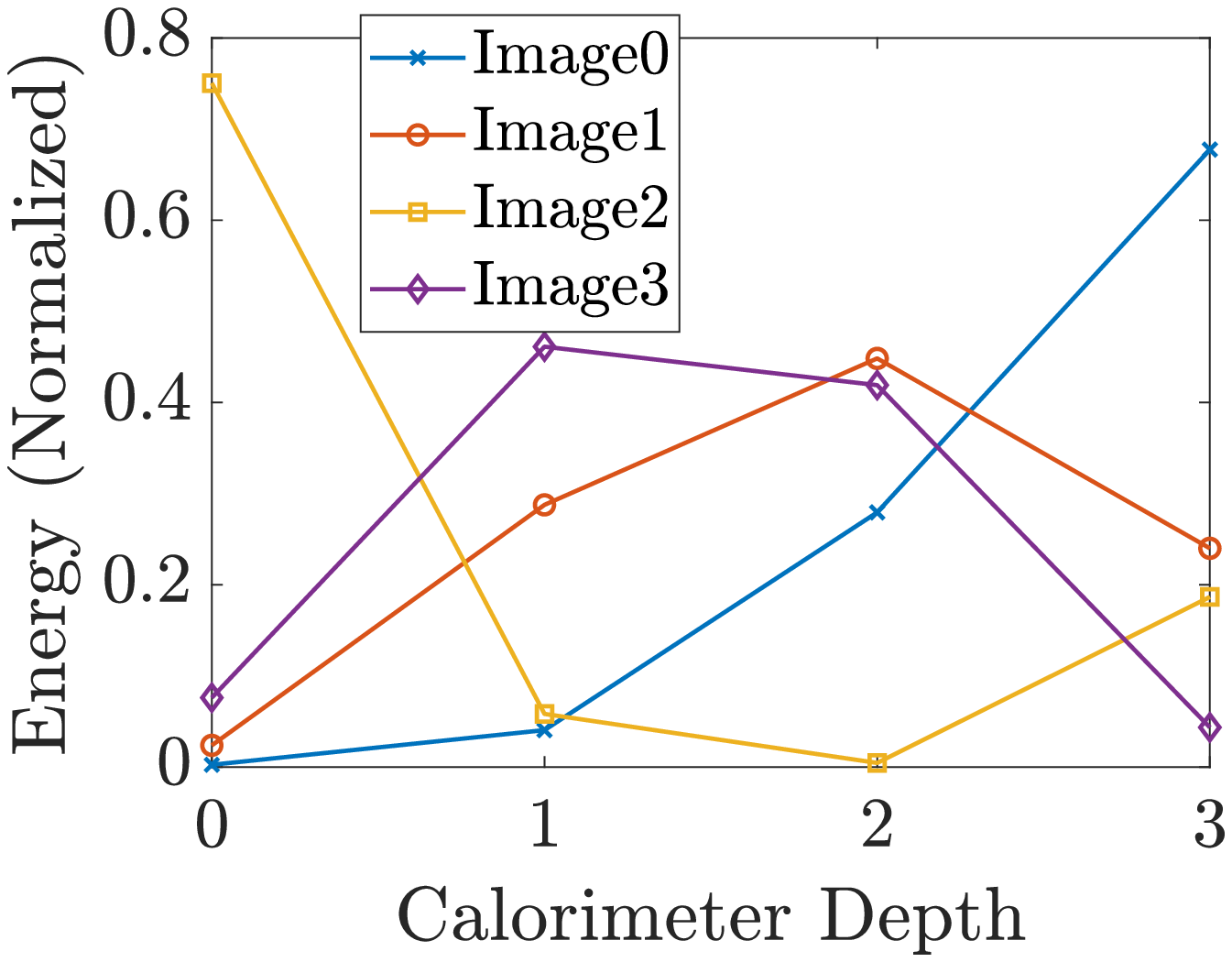}
    \caption{\em}
    \label{fig:images_N4_D2D6_opt3}
  \end{subfigure}
  \begin{subfigure}{0.245\textwidth} 
    \includegraphics[width = \textwidth]{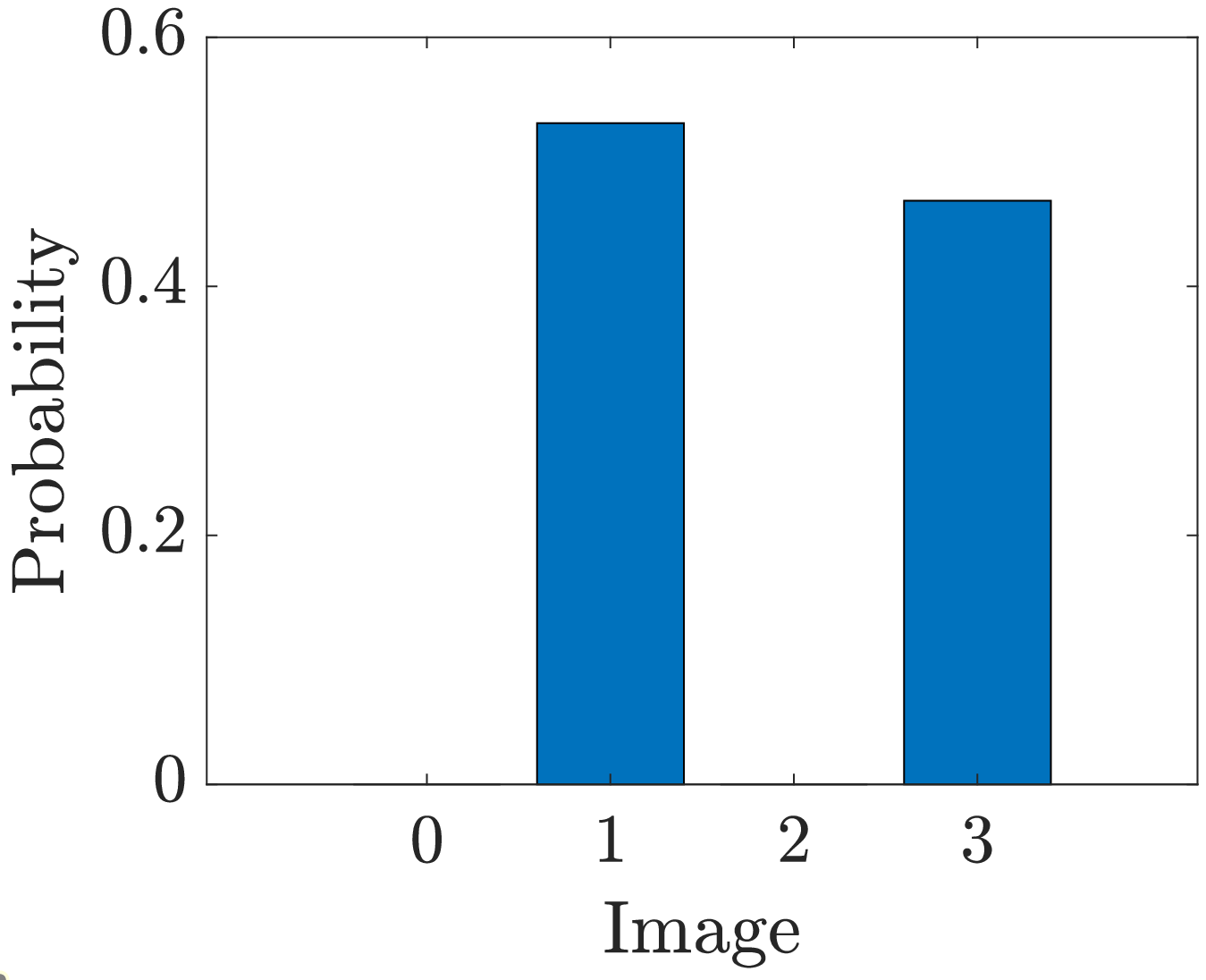}
    \caption{\em}
    \label{fig:prob_N4_D2D6_opt3}
  \end{subfigure}
  \begin{subfigure}{0.245\textwidth} 
    \includegraphics[width = \textwidth]{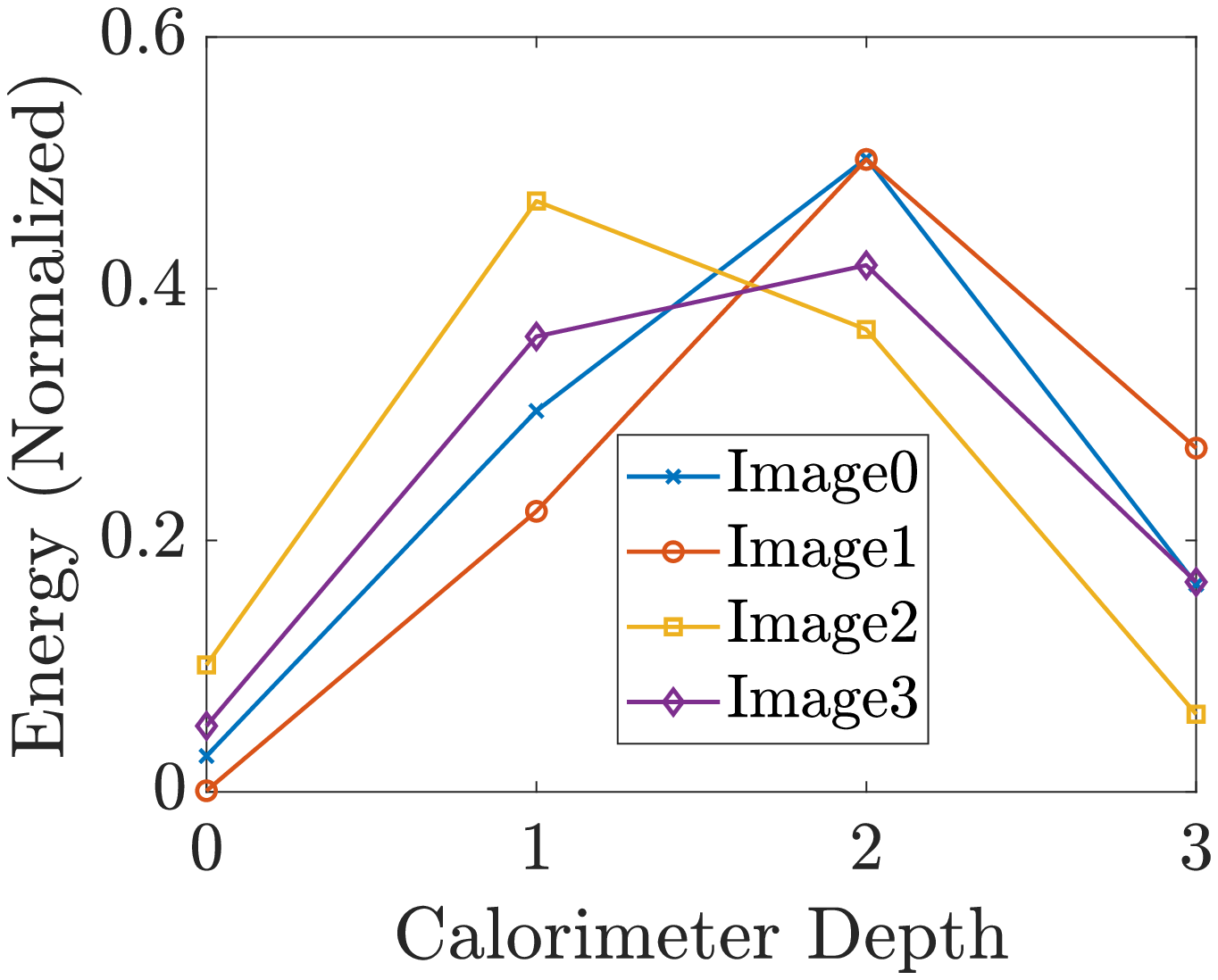}
    \caption{\em}
    \label{fig:images_N4_D2D16_opt3}
  \end{subfigure}
  \begin{subfigure}{0.245\textwidth} 
    \includegraphics[width = \textwidth]{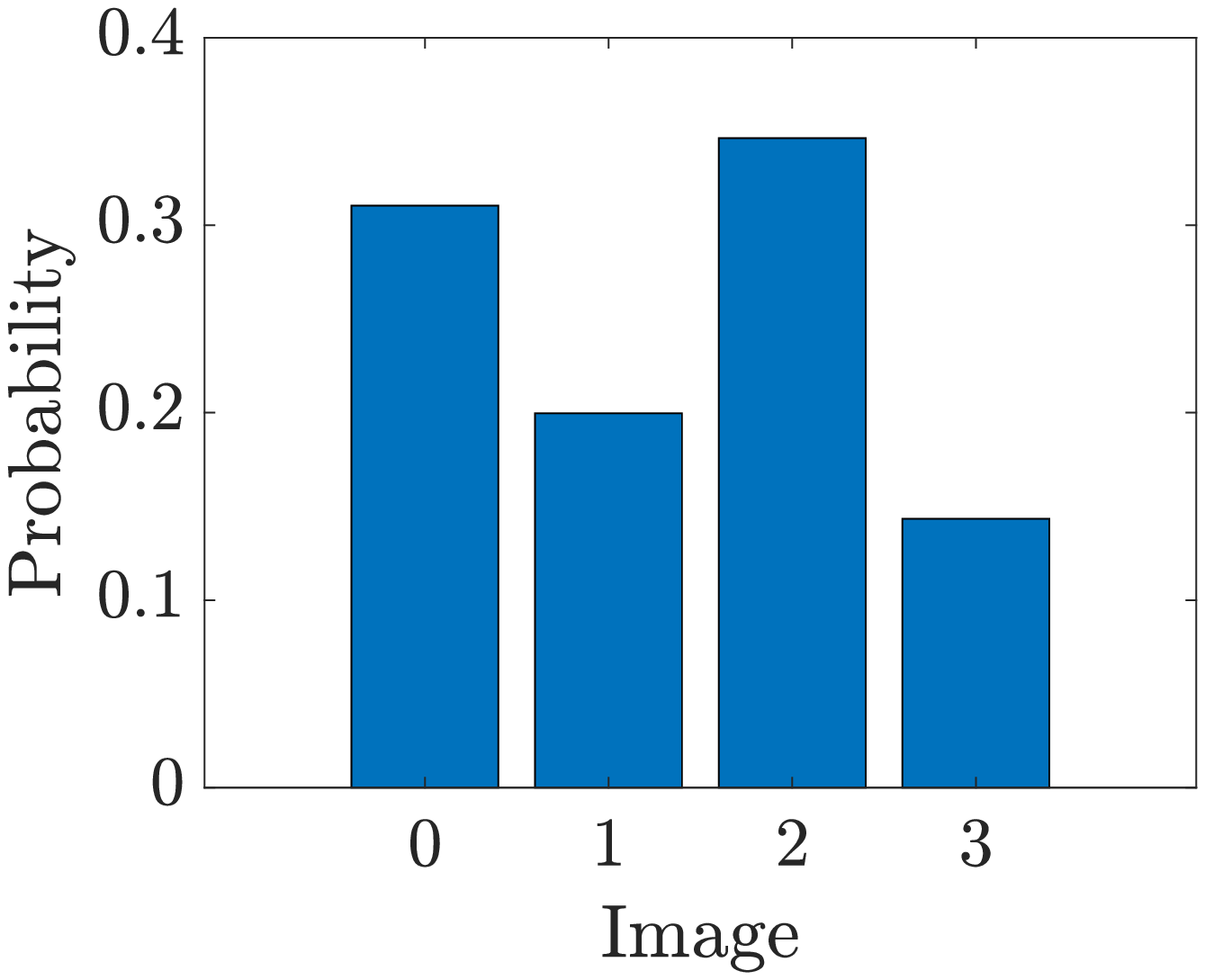}
    \caption{\em}
    \label{fig:prob_N4_D2D16_opt3}
  \end{subfigure}
  \caption{\em Images generated by PQC2 (a,c) and its distribution obtained from PQC1 (b,d) with $n = 2$, $n_1 = n_2 = 4$ and  $d_{g,1} = 2$, but with different $d_{g,2}$ :  1) $d_{g,2} = 6$ (a,b), 2) $d_{g,2} = 16$ (c,d). Although the mean images shown on \figref{fig:mean_loss_dual} are close to the real value in both cases, PQC2 with $d_{g,2} = 6$ cannot imitate correctly the real data (c.f. \figref{fig:mean_images}), while the one with $d_{g,2} = 16$ achieves to reproduce very similar images.}
  \label{fig:image_samples}
\end{figure}

As an analysis on the mean images is not enough to validate the GAN result, it is necessary to evaluate the individual images produced by PQC2 as well as the probability distribution generated by PQC1. The images generated by PQC2 with $d_{g,2} = 16$, displayed on \figref{fig:images_N4_D2D16_opt3}, are similar to the mean real images shown in \figref{fig:mean_images} with the peaks at $x = 2$ or $x = 3$. On the other hand, those generated by PQC2 with $d_{g,2} = 6$ on \figref{fig:images_N4_D2D6_opt3} contain two images, $\mathcal{I}_0$ and $\mathcal{I}_2$, which are far from the real image profile.

The probability distribution generated by PQC1 can explain this discrepancy. As shown on \figref{fig:prob_N4_D2D6_opt3}, the weights for $\mathcal{I}_0$ and $\mathcal{I}_2$ are negligible compared to those for $\mathcal{I}_1$ and $\mathcal{I}_3$, meaning that their labels, produced by the classical discriminator, are suppressed in the loss function given by \eqnref{eq:lg}. Therefore, the mean images and the  generator loss could converge to the correct value, despite considerable errors in the produced images themselves. 
Contrarily, in case of $d_{g,2} = 16$, the probability on \figref{fig:prob_N4_D2D16_opt3} implies that all 4 images have non-negligible weights, thus leading to consistency between the quality for the mean and individual images. This result certainly highlights the importance of choosing a correct structure of dual-PQC GAN model in order to prevent any bias during training. 

\begin{figure}[h]
  \centering
  \begin{subfigure}{0.38\textwidth} 
    \includegraphics[width = \textwidth]{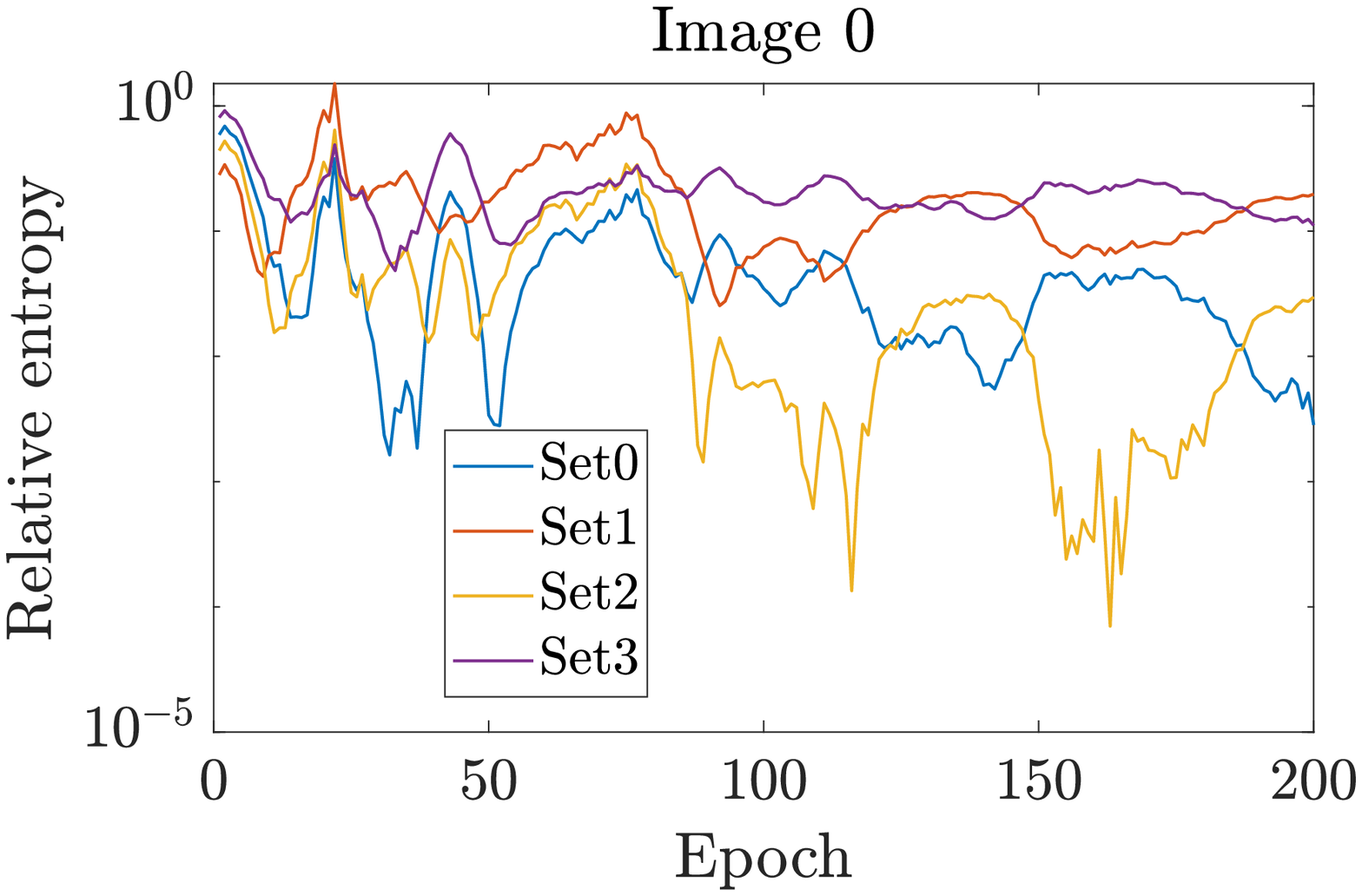}
%     \caption{\em Image 0.}
  \end{subfigure}
  \begin{subfigure}{0.38\textwidth} 
    \includegraphics[width = \textwidth]{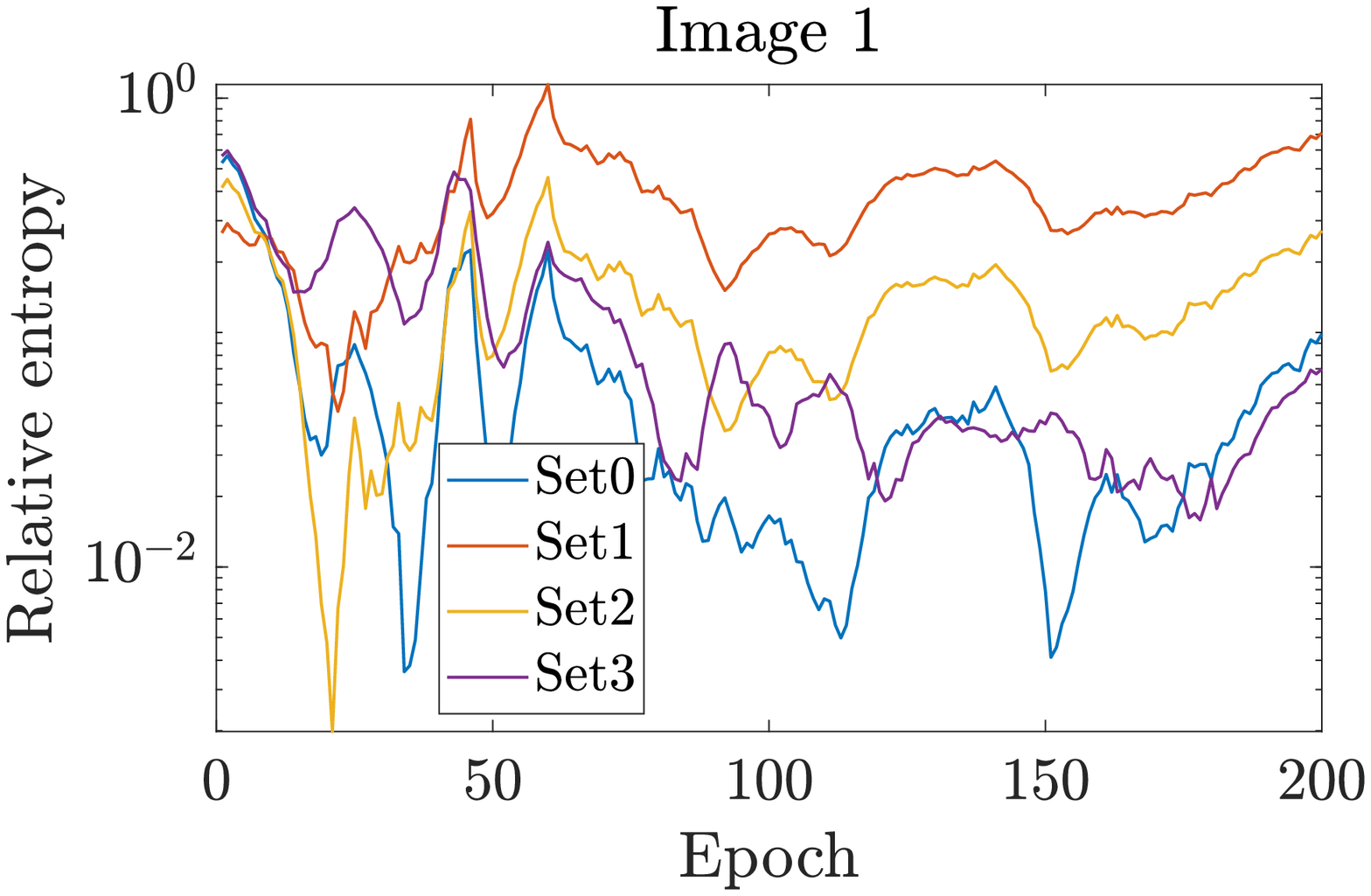}
%     \caption{\em Image 1.}
  \end{subfigure}
  \begin{subfigure}{0.38\textwidth} 
    \includegraphics[width = \textwidth]{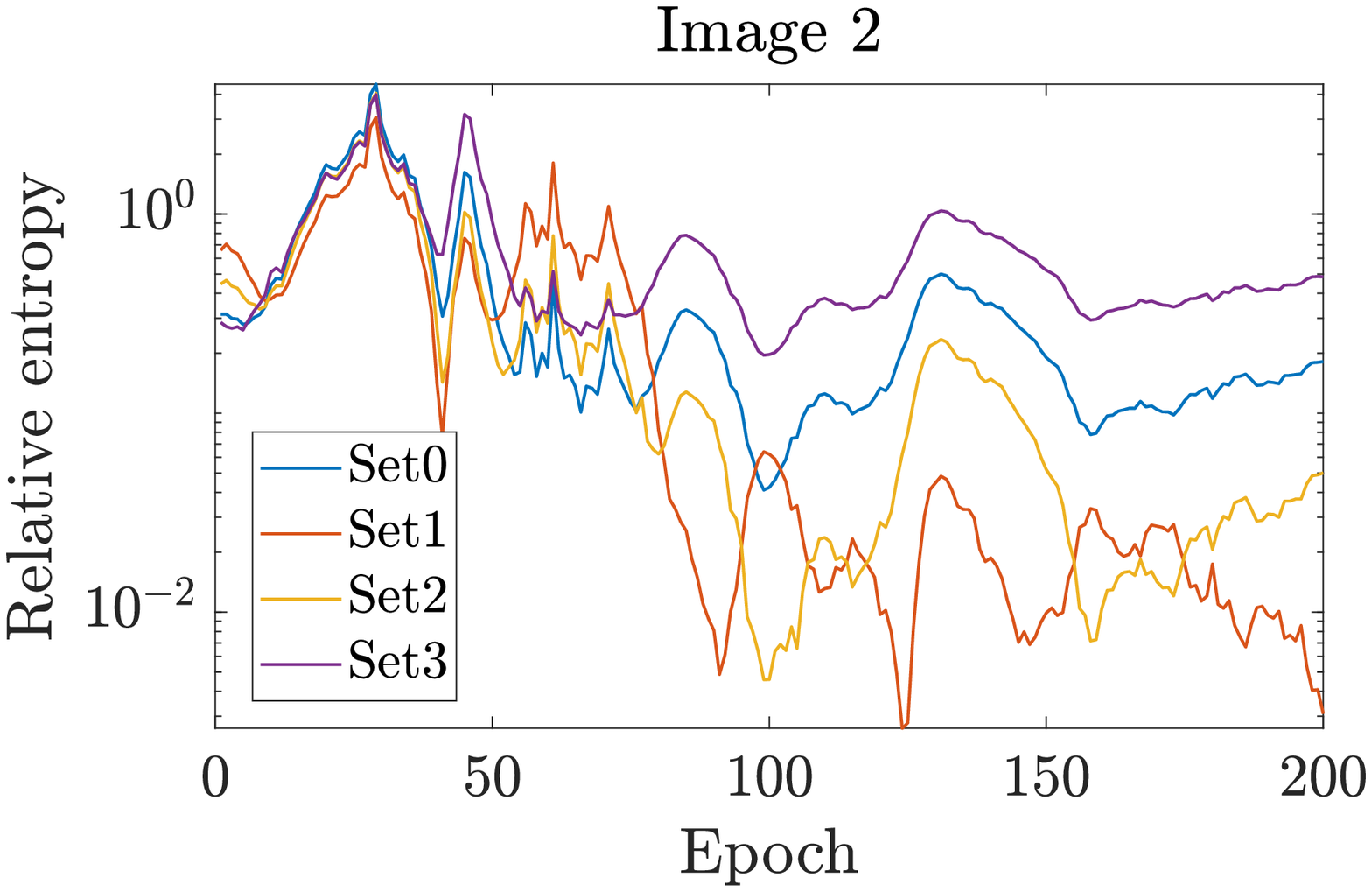}
%     \caption{\em Image 2.}
  \end{subfigure}
  \begin{subfigure}{0.38\textwidth} 
    \includegraphics[width = \textwidth]{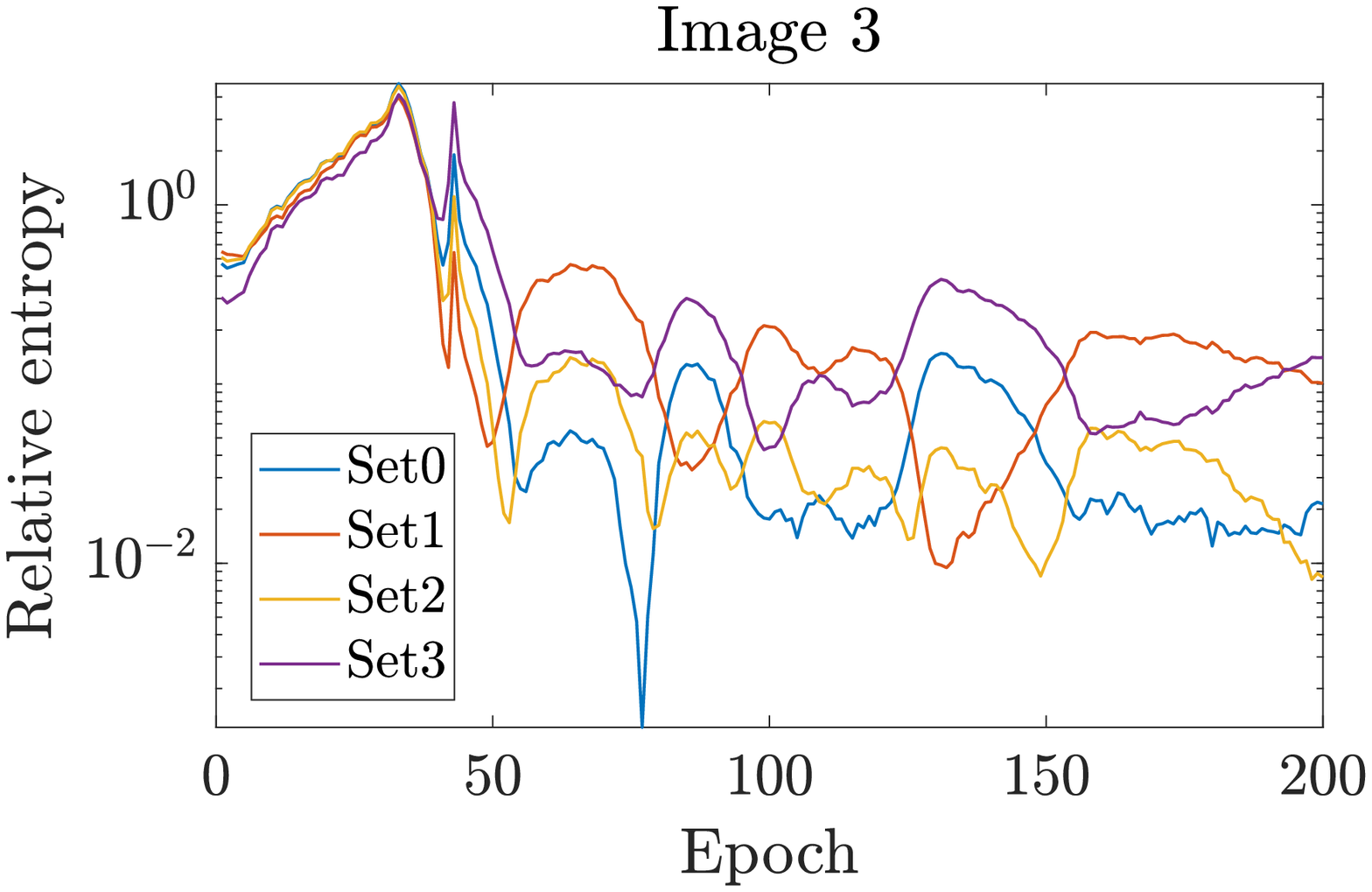}
%     \caption{\em Image 3.}
  \end{subfigure}
  \caption{\em Relative entropy of generated images $\mathcal{I}_0,..., \mathcal{I}_3$ with respect to average of real image classes, shown on \figref{fig:mean_images}. The images are generated via dual-PQC GAN with $n = 2$, $n_1 = n_2 = 4$, $d_{g,1} = 2$ and $d_{g,2} = 16$. According to minimum relative entropy at the end of the training, one-to-one correspondence can be established between real and generated images.}
  \label{fig:individual_rel_entr}
\end{figure}

Finally, the quality of individual images is evaluated by calculating the relative energy between the real images Set~$i$, shown on \figref{fig:mean_images}, and the generated images $\mathcal{I}_i$ for $i = 0,1,2,3$, as displayed on \figref{fig:individual_rel_entr}. Considering the lowest relative entropy values across the last 4 epochs (200 epochs are run in total), a bijection between real images and generated images can be constructed : $\mathcal{I}_0 \to$ Set0, $\mathcal{I}_1 \to$ Set3, $\mathcal{I}_2 \to$ Set1 and $\mathcal{I}_3 \to$ Set2. This result gives a quantitative proof that PQC2 can reproduce four different sets of images in the real training data. Despite this optimistic affirmation, the relative entropy large instability is the main point that should be improved in future studies. 

Unfortunately, the number of qubits in the quantum generator scales not only with the number of pixels in one image but also with the number of images that can be produced, while in original GAN, the system size mainly scales with image size. This fact represents the largest limitation of the model and it requires further improvement. We are currently investigating a solution feeding  extra noise to the remaining PQC2 $n_2 - n_1$ qubits, while PQC1 keeps generating a probability distribution over the zero-noise image.

\section{Conclusion}
\label{sec:conclusion}

This work presents a dual-PQC GAN model, a prototype of quantum GAN with two quantum generators, sharing the role of a single generator. One of the generators is responsible for reproducing the distribution over images, while the other for building amplitude distributions over pixels on a single image. If we supplement the input of PQC2 with a sample from PQC1 and some (continuous) noise, then we can see that PQC1 really is a distribution over means of different classes of images, and the noise allows generation of typical samples from the class. The results obtained prove that this model can generate individual image samples and their probability distribution, similar to the training set. 

It is also worth noting that, as the number of possible images (or classes of images if noise is added) grows \textit{exponentially} with $n_1$, the fact that we sample from a finite set of images (or classes of images if noise is added) rather than a continuum of typical images (as in the corresponding classical case) is unlikely to seriously compromise performance.

An interesting question for future research would be how to reproduce an arbitrary number of outputs with the dual-PQC GAN model.  One possible way is to introduce a \textit{complex-valued} noise to the remaining qubits in the PQC2. Using a fixed sampler instead of a trained quantum circuit for PQC1 to pass an entangled quantum state to PQC2 is another possible approach. Throughout future studies, we look forward to building a more advanced quantum GAN model to imitate the performance of classical GAN.

% Apart from the text (above) giving a bit more commentary about the development of the dual PQC GAN idea, I think the main ommision is a discussion about the significance of the fact that PQC1 generates a probability distribution over a finite set of images, whereas a classical GAN samples from a continuum. I am not entirely sure where this discussion should go, but I think we should make the following two points (the latter already partially made, but could be strengthened):
% As the number of images grows as $2^{n_1}$ we can make the point that, as the problem size (ie size of the PQCs) the exponential growth makes this ``almost infinite'' -- in practise it is unlikely that we will sample the same image multiple times, and the co-training of the two PQCs will have to achieve the fundamental goal (of classical and quantum machine learning) of generalising to unseen data.
%  If we supplement the input of PQC2 with a sample from PQC1 and some (continuous) noise, then we can see that PQC1 really is a distribution over means of different classes of images, and the noise allows generation of typical samples from the class.

%
% BibTeX or Biber users please use (the style is already called in the class, ensure that the "woc.bst" style is in your local directory)
\bibliography{reference_abbr.bib}

\end{document}